\newcommand{\gpip}{\displaystyle{{\bar g^2/2\pi}}}

\newcommand{\eq}{\begin{equation}}
\newcommand{\en}{\end{equation}}
\newcommand{\BEF}{\begin{figure}}\newcommand{\EF}{\end{figure}}
\newcommand{\bea}{\begin{eqnarray}}
\newcommand{\eea}{\end{eqnarray}}

\newcommand{\J}{\J_{\mu ab}}
\newcommand{\T}{\T^{\mu\nu}}

\newcommand{\unity}{1\kern-.65mm \mbox{\form l}}
\newfont{\form}{cmss10}

\newcounter{lett2}

\documentstyle[prl,aps,12pt]{revtex}

\begin{document}
\draft
\title{\Large\bf Partition functions of chiral gauge theories on the two 
dimensional torus and their duality properties\ }
\author{L. Griguolo$^{(1,2)}$ and D. Seminara$^{(3)}$}
\address{\it  $^{(1)}$ Center for Theoretical Physics, Laboratory
                         for  Nuclear Science,\\
                         Massachusetts Institute of Technology, Cambridge,
                         Massachusetts 02139, USA.\\
              $^{(2)}$ Dipartimento di Fisica "G.Galilei", 
                      INFN, Sezione di Padova,\\
                      Via Marzolo 8, 35131 Padua, Italy\\
              $^{(3)}$ Department of  Physics, Brandeis University,
                        Waltham, MA 02254, USA\\}           
\date{Received \today}
\maketitle
\medskip
\font\ninerm = cmr9
\pagestyle{empty}
\begin{abstract}
\ninerm\noindent
Two different families of abelian chiral gauge theories on the 
torus are investigated: the aim is to test the consistency of 
two-dimensional anomalous gauge theories
in the presence of global degrees of freedom for the gauge field. An explicit
computation of the  partition  functions shows that unitarity is recovered
in particular regions of parameter space and that the effective dynamics
is described in terms of  fermionic interacting models. For the first
family, this connection with fermionic models uncovers an exact duality
which is conjectured to hold in the nonabelian case as well.
\end{abstract}
\vfill
\begin{flushright}
\end{flushright}
\newpage
\pagestyle{plain}

Chiral generalizations of the Schwinger model
\cite{Jac85,Ro86,Gri94} have given the possibility to test the
non--perturbative dynamics of gauge theories in the presence of local
anomalies. 
It was shown that a consistent theory emerges in spite of the fact that 
order by order in perturbation theory the anomaly breaks the unitarity: 
unfortunately a similar result is not available in four dimensions, where
only 
perturbative calculations are possible. The physics of anomalous gauge 
theories appears to depend, in two dimensions, on a real parameter, 
called the Jackiw-Rajaraman 
parameter, reflecting the regularization ambiguity and usually denoted 
by $a$: in particular (in the pure chiral case) unitarity requires $a>1$. 
The spectrum of the theory, its vacuum structure and the nature of 
fermionic states has been obtained in \cite{Jac85,Ro86} for the pure 
chiral case and generalized to mixed vector--axial 
couplings in \cite{Gri94}, where a precise relation with the massless 
Thirring model has been pointed out (see also \cite{Belve} for a recent 
discussion). 
Related investigations have been also carried out in string theory: 
there the possibility of exploring the anomalous dynamics induced by 
Weyl \cite{Weyl} and Lorentz \cite{Pery92,Basti92} anomaly has triggered 
the attention of many researchers. In the context of string theory this 
is not, however, the only reason of interest. In fact gauge fields 
interacting with chiral world--sheet 
fermions on arbitrary Riemann surface were examined in 
\cite{Tombo89}, providing a mechanism for spontaneous symmetry 
breaking.

\noindent
In this letter we present the exact partition function for two 
different families of (abelian) gauge theories on the torus $T^2$, 
taking anti-periodic boundary conditions 
for fermions. The aim is to investigate the consistency and the 
behavior of an anomalous gauge theory in presence of the global degrees of 
freedom of the gauge field, linked to the non--trivial homology cycles 
of the underlying manifold. Large gauge invariance comes therefore into 
play and it is interesting to test the unitarity in this more general 
context. We notice that our results have a direct interpretation in 
finite temperature field theory, when the flat limit is taken and the 
$x$ direction is decompactified. The computations we present can be 
extended to Riemann surface of any genus with minimal technical complications, 
however all the relevant features are already present 
in the genus-one calculation, due to the non self--adjoint character of the 
Dirac--Weyl operator and to the presence of the harmonic piece in the Hodge 
decomposition of the gauge field. A careful application of $\zeta$--function 
regularization allows us to obtain the relevant functional determinants 
without any analityc continuation on the chiral couplings 
(at variance with\cite{Wi95}) or modular invariance requirements 
\cite{Vafa}, leading to a result expressed in term of theta-function 
with characteristic.

\noindent
The first model we consider is the generalized  chiral Schwinger model:
\begin{equation}
\label{Lagrangian}
{\cal L}\!=\! i \bar \psi\gamma^\mu \!\left (\!\nabla_\mu +
i \frac{1+ r \gamma_5}{2} A_\mu \!\right )\psi\!
-\!\frac{g^2}{4}\bar\psi\gamma^\mu\psi \bar\psi\gamma_\mu\psi
\!-\!\frac{1}{4 e^2}F_{\mu\nu}F^{\mu\nu}.
\end{equation}
The geometry is described by the zweibein $e^a_\mu$ 
($g_{\mu\nu}=e^a_\mu e_{a\mu}$) that, with a suitable choice of 
the  Lorentz frame, can always been written in the form
\begin{equation}
e^a_\mu=e^{\sigma(x)} \hat e^a_\mu=e^{\sigma(x)}
\left (
\begin{array}{cc}
\tau_2 & \tau_1\\
0 & 1
\end{array}
\right),
\end{equation}
the index $a$ spans the columns, while the index $\mu$ runs over the rows. 
The exponent $\sigma(x)$ is the conformal factor, $\tau=\tau_1+i\tau_2$ 
is the Teichm\"uller parameter and the fundamental region for 
the coordinates has been taken 
to be the square $0\le x <L$  and $0\le y <L$.  The covariant derivative 
for the Dirac spinor is $\displaystyle{\nabla_\mu\equiv \partial_\mu
-i\frac{\gamma_5}{2}\omega_\mu}$, the corresponding spin-connection 
is then computed from the condition of vanishing torsion
$\displaystyle{
\omega_\mu=-\frac{\hat g_{\mu\nu}}{\tau_2}\epsilon^{\rho\nu}
\partial_\rho \sigma.}
$
The gamma matrices $\gamma^\mu$ in curved space-time are related to 
the flat ones as  $\gamma^a e^\mu_a$\footnote{Our notations, regarding
the gamma matrix and the $\epsilon_{\mu\nu}$ 
tensor are $\epsilon^{01}=\epsilon_{01}=1, \ \ \{\gamma_a, \gamma_b\}=
2\delta_{ab},\ \   \gamma_a \gamma_5=i \epsilon_{ab}\gamma^b $.}. 
The fermions appearing in the Lagrangian (\ref{Lagrangian}) are
chosen to satisfy anti-periodic boundary conditions:
$\psi(x+L,y)=-\psi(x,y)$ and $\psi(x,y+L)=-\psi(x,y).$
The quantity $r$ is a real parameter interpolating between the vector
($r=0$) and the completely chiral ($r=\pm 1$) Schwinger model. The 
Thirring-like interaction, governed by the coupling constant $g^2$, 
has been introduced in order to simplify the analysis of the final
result. 
The gauge field $A_\mu$ is taken 
to be on a trivial $U(1)-$bundle. At classical level this Lagrangian is 
invariant under the local transformations
\begin{equation}
A^{\prime}_{\mu}\frac{1+ r\gamma_5}{2}=
A_{\mu}\frac{1+r\gamma_5}{2}+iU^{-1}\partial_\mu U,\ \ \ \ \
\psi^{\prime}=U\psi,
\end{equation}
with $\displaystyle{ U=\exp[2\pi 
i(\frac{1+r\gamma_{5}}{2})\Lambda]}$.
Actually being on the torus we have to impose that the gauge transformation 
$U$ be well-defined, namely $U(x+L,y)=U(x,y)$  and  $U(x,y+L)=U(x,y)$.
While for general $r$ a periodic $U$ entails a periodic $\Lambda$,
for rational $r$ ($\displaystyle{=\frac{p}{q}}$ with $p$ and $q$ relative
prime) we have only
\begin{equation}
\Lambda(x+L,y)-\Lambda(x,y)= q~ n_1\ \ \   
\Lambda(x,y+L)-\Lambda(x,y)= q~ n_2\ \ \ \ \ \  {\rm with} \ \ 
n_1,n_2\in Z\!\!\!Z.
\end{equation}
[As a matter of fact, the rational nature of  $r$ allows the existence of
large gauge transformations.] Despite this rich classical structure, the 
quantum theory, for $r\ne 0$, will always exhibit a gauge anomaly that 
potentially undermines its consistency. However the following evaluation
of the partition function ${\cal Z}$  will show that unitarity can be 
recovered in particular  regions of the parameter space. 
A path integral expression for ${\cal Z}$ is given by
\begin{equation}
\label{path}
{\cal Z}=\int {\cal D}\bar \psi {\cal D}\psi {\cal D} A_\mu
{\cal D} B_\mu {\rm exp}\left (-\int d^2x  \sqrt{g} {\cal L}_c\right),
\end{equation}
where ${\cal L}_c={\cal L}+g(\bar\psi\gamma_\mu\psi)B_\mu +B_\mu B^\mu$
and $B_\mu$ is the ``usual'' auxiliary field disentangling the 
current-current interaction. The fermionic integration  is now reduced 
to compute the determinant of the Dirac-Weyl operator
\begin{equation}
\label{bubu}
D_{DW}=i \gamma^\mu \left(\nabla_\mu +
i \frac{1+r \gamma_5}{2} A_\mu -i g B_\mu\right ).
\end{equation}
To this purpose, we consider the Hodge decomposition for the one-forms 
$A_{\mu}$ and $B_{\mu}$:
\begin{equation}
\label{Hodge}
A_{\mu}=\partial_\mu 
\phi_{1}-\eta_{\mu}^{~\nu}\partial_{\nu}\phi_{2}+\frac{2\pi}{L}a_{\mu},
\ \ \ \
B_{\mu}=\partial_\mu 
\chi_{1}-\eta_{\mu}^{~\nu}\partial_{\nu}\chi_{2}+\frac{2\pi}{L}b_{\mu},
\end{equation}
where $\eta_{\mu\nu}\equiv \sqrt{g}\epsilon_{\mu\nu}$ is the usual
volume  two-form and $a_{\mu}$ and $b_{\mu}$ are two harmonic fields
satisfying the equations
$\nabla_\mu a^\mu=\eta^{\mu\nu}\nabla_{\mu}a_{\nu}=0$ and 
$\nabla_\mu b^\mu=\eta^{\mu\nu}\nabla_{\mu}b_{\nu}=0$. 

\noindent
With the help of eqs. (\ref{Hodge}) the Dirac-Weyl operator in 
(\ref{bubu}) can be now cast  in the form
\begin{equation}
{D}_{DW}=
\exp\left[- \frac{i}{2} F-\frac{3}{2} \sigma +
\frac{\gamma_5}{2} G  \right]~{\cal D}^G_{DW}~
\exp\left[ \frac{i}{2} F+\frac{\sigma }{2}+
\frac{\gamma_5}{2} G\right],
\end{equation}
where
$F=\left ( \phi_1 + i r \phi_2 -2 g \chi_1 \right )$
and $G=\left (\phi_2-i r\phi_1 -2 g \chi_2 \right)$.
The operator ${\cal D}^G_{DW}$ depends only on the global modes 
of the vector fields
\begin{equation}
{\cal D}^G_{DW}=i \gamma^\mu
\left [\partial_\mu+\frac{2\pi}{L}\left
(\frac{a_\mu}{2}- g b_\mu \right )+
\frac{2\pi}{L}\hat \eta_{\mu}^{~\nu}(\frac{r}{2}a_\nu)\right ],
\end{equation}
By means of the $\zeta$--function technique \cite{Hawk}, 
the contribution of $F$, $G$ and  $\sigma$ to the determinant 
of  ${D}_{DW}$ can be factorized out, giving as result:
\begin{equation}
\label{factorization}
\det (D_{DW})= \det ({\cal D}^G_{DW})\exp \left
(-S_{Loc.}(\phi,\chi,\sigma)\right )   
\end{equation}
$S_{Loc.}$ contains two  parts: the Liouville action generated by the 
Weyl anomaly and  the 
abelian {\it WZWN}  term arising from the gauge anomaly. 
Explicitly it reads
\begin{eqnarray}
&& S_{Loc.}(\phi,\chi,\sigma)=\frac{1}{96\pi}
\int dx^2 \sqrt{\hat g}\left (\hat g^{\mu\nu}
\partial_\mu\sigma\partial_\nu\sigma+\lambda e^{2\sigma}
\right)+\nonumber\\
&&\!\!-\!\!\ \frac{1}{8\pi}
\int dx^2 \sqrt{ g}(\phi_1- i r\phi_2 -2 g \chi_2 )\triangle
(\phi_1- i r\phi_2 -2 g \chi_2).
\end{eqnarray}
However, due to the loss of gauge invariance, an intrinsic ambiguity 
is present in the factorization  (\ref{factorization}).
In fact changing our regularization scheme in a suitable way, we
can generate a four-parameter family of local counterterms, which
might be added to $S_{Loc.}$
\begin{equation}
\frac{1}{8\pi}\int dx^2 \sqrt{\hat g}\hat g^{\mu\nu}( a A_\mu  A_\nu +
b B_\mu  A_\nu + c B_\mu  B_\nu +
d \hat\eta^{\mu\nu}A_\mu  B_\nu ).
\end{equation}
Actually, three of these parameters ($b$, $c$ and $d$) can be
ignored because they simply correspond to a  trivial
rescaling of the coupling constants $g$, $e$ and $r$.

\noindent
After changing integration variables from $A_\mu$ and $B_\mu $ to 
$\phi_i,\ \chi_i,\ a_\mu$ and $b_\mu$, the initial path
integral in eq. (\ref{path}) reads ${\cal Z}\equiv {\cal Z}_{Loc.}
\times {\cal Z}_{Glob.}$, with
\begin{eqnarray}
\label{path1}
&&{\cal Z}_{Loc.} =4\pi^2 {\det}^\prime (\triangle)^{-2}
\int {\cal D} \vec{\Phi}^\prime (x)
{\rm exp}\left (-S_{Liou.}(\sigma)-\frac{1}{2}
\int d^2x\sqrt{g}\vec{\Phi}(x)^t A
\vec{\Phi}(x)\right )
\nonumber\\
&&{\cal Z}_{Glob.}=\int^{\infty}_{-\infty} da_\mu db_\mu
\exp\left (-\frac{a\pi}{2}\sqrt{\hat g}\hat g^{\mu\nu}a_\mu a_\nu
-4\pi^2\sqrt{\hat g}\hat g^{\mu\nu}b_\mu b_\nu\right)
\det ({\cal D}^G_{DW}).
\end{eqnarray}
We have introduced a vector notation $\vec{\Phi}\equiv(\phi_1, 
\phi_2,\chi_2, \chi_1)$ and a fourth  order differential matrix 
operator $A$ that can be easily read from $S_{Loc}$ and the Maxwell term.
%
%
The prime on the measure means that the functional integration must 
be carried out  only over the non constant modes of $\chi_i$ and 
$\phi_i$ and $4\pi^2\det^\prime(\triangle)^{-2}\,\,\,$\footnote{As usual 
the 
symbol $\det^\prime$ means that the zero eigenvalue is excluded.}
is  the jacobian of the change of variables. The integration over 
$a_\mu$ and $b_\mu$ is extended over $(-\infty, +\infty )$. This 
deserves an explanation: When $r$ is rational, due to presence of 
large gauge transformations, $a_\mu$ and $a_\mu+qn_\mu$ are gauge equivalent 
$(n_\mu=(n_1, n_2), n_{1,2}\in Z\!\!\!Z)$ and the factorization of the gauge 
volume would lead to the integration over the fundamental domain $[0,q]$. 
However there is no reason to perform this factorization since the invariance 
is broken by the anomaly; therefore we extend the integration over all the 
connections space and not limit ourselves to the gauge orbits. 
${\cal Z}_{Loc.}$ is easily computed performing a standard Gaussian
integration 
and gives
\begin{equation}
\label{Poiu}
{\cal Z}_{Loc.} =2\sqrt{a\bar{a}}
\det\left (-\frac{\triangle}{M^2} +1\right 
)^{-1/2}\exp\left[-S_{Liou.}\right],\ \  {\rm with}\ \
M^2\equiv  \frac{\bar e^2}{4\pi}
\frac{\bar a^2}{\bar a - 1}.
\end{equation}
Here $\bar a=a \displaystyle{\left (1+\frac{g^2}{2\pi}+\frac{1-r^2}{a}
\right )}$ and
$\bar e^2=\displaystyle{{e^2}/{\displaystyle{\left(1+\frac{g^2}{2\pi}+
\frac{1-r^2}{a}\right )}}}$. Even though the 
determinant (\ref{Poiu}) cannot be computed explicitly in a generic metric
background,  its  physical interpretation 
is clear. It represents the partition function of a massive boson with a
mass given by $M$. Particular care must be paid to the singular case 
$\hat a=r^2$; in fact  $M^2\to\infty$ for  this choice of the parameters. If 
we  carefully recalculate  ${\cal Z}_{Loc.}$ in this limit, we get $1$,
namely
the massive degree of freedom has decoupled from the physical spectrum.

\noindent
The evaluation of ${\cal Z}_{Glob.}$ is more involved, requiring an 
explicit $\zeta-$function calculation. The determinant of ${\cal D}_{DW}^G$
is evaluated as the square root of $\det({\cal D}_{DW}^G)^2$  can be 
by means of the well-known relation
\begin{equation}
{\rm det}[ D_{DW}^G ]=\exp\left[-1/2\frac{d}{d s}\zeta_{D_{DW}^G}(s)
\right]_{s=0}.
\end{equation}
[No relevant
ambiguity appears in $\zeta-$function formalism in even dimensions as 
shown in\cite{NOI}]. Imposing the anti-periodic boundary conditions and 
solving the eigenvalue problem we get the $\zeta-$function
\begin{equation}
\zeta(s)=2\left(\frac{L\tau_2}{2\pi}\right )^{2s}\!\!\sum_{ Z^2}
\left [\left (n_1+H_1+\frac{1}{2}-\tau_1\left ( n_2+H_2+\frac{1}{2}
\right )\right )^2\!\!\!+\tau_2^2  \left (n_2+H_2+\frac{1}{2}\right )^2
\right  ]^{-s}\!\!\!\!,
\end{equation}
where the symbol $H_\mu$ stands for the combination
$\displaystyle{
H_\mu=\frac{a_\mu}{2}- g b_\mu -\frac{r}{2}\eta_{\mu}^{~\nu}a_\nu .}$
The computation of $\zeta^\prime (0)$ is 
quite technical and it makes use of Poisson resummation and analytic 
continuation in $s$. 
The final result, which can be expressed in terms of 
theta functions, is
\begin{eqnarray}
&&\!\!\!\!\!\!\!\!\!\!\!\!
\zeta^\prime(0)\!=\!
-\pi r^2\sqrt{\hat g} \hat g^{\mu\nu}\!\!a_\mu a_\nu \!\!
-4\pi i r a_2\Biggl (\frac{a_1}{2}-g b_1 \Biggr )\\
&&\!\!\!\!\!\!\!\!\!\!\!
-2\log\!\!\left( \frac{1}{|\eta(\tau)|^2}\Theta \left[\!\!
\begin{array}{c}
\displaystyle{g b_2-\frac{r+1}{2} a_2}\\
\displaystyle{-g b_1+\frac{r+1}{2} a_1}\\
\end{array}
\!\!\right]\!\!(0,\tau)\
\Theta^* \left[\!\!
\begin{array}{c}
\displaystyle{g b_2+\frac{r-1}{2} a_2} \\
\displaystyle{-g b_1+\frac{1-r}{2} a_1}\\
\end{array}
\!\!\right]\!\!(0,\tau)\right ) \nonumber
\end{eqnarray}
To perform the integration over the 
flat connections $a_\mu$ and $b_\mu$, one has to expand the 
$\theta-$functions in their series representation, integrate term by term 
and finally resum the ensuing series. This straightforward, but tedious 
exercise leads to
\begin{equation} 
{\cal Z}_{Glob.}={1\over 
2\sqrt{a\bar{a}}}\frac{1}{|\eta(\tau)|^2}
\Theta(0,\Lambda(\tau,\bar{g})),
\end{equation}
where $\Theta(0,\Lambda(\tau,\bar{g}))$ is a theta-function with
characteristic,
whose   covariance matrix $\Lambda(\tau,\bar{g})$ is
\begin{equation}
\label{covariance}
\Lambda(\tau,\bar{g})=\left (\begin{array}{cc}
\tau &  0\\
0 & -\bar{\tau}\end{array} \right )+
i {\tau_2 \displaystyle{\bar{g}^2\over 2\pi}\over 2\left(1+
\displaystyle{\bar{g}^2\over 2\pi}\right)}\left 
(\begin{array}{cc}
\displaystyle{\bar{g}^2\over 2\pi} &  -2-
\displaystyle{\bar{g}^2\over 2\pi}\\
-2-\displaystyle{\bar{g}^2\over 2\pi} &
\displaystyle{\bar{g}^2\over 2\pi}\end{array} \right ),
\end{equation}
and $
\Theta(0,\Lambda)=\sum_{\vec{n}\in Z^2}\exp[i\pi \vec{n}\,\Lambda\, \vec{n}].
$
The parameter $\displaystyle{\bar{g}^2\over 2\pi}$ is defined as
$\displaystyle{{\bar{g}^2\over 2\pi}={{g}^2\over 2\pi}+{1-r^2\over a}}.$
This is (apart from the prefactor that is cancelled by the analogous one 
in ${\cal Z}_{Loc.}$) the partition function on the torus of the abelian 
Thirring model of coupling constant $\bar{g}^2$. This is not a surprise, 
in fact in \cite{Gri94} it was shown on the plane at level both of operators 
and correlation functions, that the generalized chiral Schwinger model is 
equivalent to a massive boson plus an effective Thirring interaction. 
There ($g^2=0$) the effective Thirring coupling was
$\displaystyle{\bar{g}^2\over 2\pi}=\displaystyle{1-r^2\over a}.$
Here we see that the addition of a bare Thirring interaction simply 
leads to a theory in which the couplings sum. More remarkably we 
notice that the careful treatment of the global degrees of freedom 
allowed us to reproduce the plane behaviour.

\noindent 
The final expression for the partition function turns out to be
\begin{equation}
{\cal Z}=\exp
[-S_{Liou.}]{\rm det}\left(-{\Delta\over M^2}+1\right)^{-{1\over 2}}
\Theta(0,\Lambda(\tau,\bar{g})).
\label{parto}
\end{equation}
In ref. \cite{Adamo} it has been argued that a similar 
factorization in a massive and in a conformal invariant sector holds 
for all two dimensional gauge theories. 

We are ready now to discuss the unitarity properties of the model. 
It is well known that unitarity requires for the Thirring theory that 
${\bar{g}^2\over 2\pi}>-1$: this fact is easily understood as singularities 
appear in 
$\Theta$-function. Next the mass must be real therefore $\bar a>1$. 
In absence of bare Thirring interaction $(g^2=0)$ these constraints are 
equivalent to the same unitarity window found on the plane \cite{Gri94} 
${\it i.e}$ $a>r^2$. In the form written here the condition on $\bar a$ is 
exactly the same as for $a$ in the pure chiral Schwinger model 
\cite{Jac85}, to which ours reduces when $\bar{g}=0$ (the 
fermionic part of the partition function collapses, in that case, to the 
one of free fermions). The novelty of the interpolating coupling $r$ 
consisting essentially into a rescaling of the electric charge 
($e\rightarrow \bar{e}$) and the generation of a current-current 
interaction term. We notice that the fermionic part is conformal 
invariant as the abelian Thirring model is, while only in the (singular) 
limit $\bar a\rightarrow 1$ (were apparently the boson mass diverges) we 
recover (by decoupling) the same in the bosonic sector. 

\noindent
A further remark is that many different choices of the initial parameter 
($r,g,a$) leads to same partition function: let us consider the case 
$g^2=0$. Giving $\displaystyle{\frac{\bar g^2}{2\pi}}$ and $M^2$ we have
in general 
two values ($a_\pm, r_\pm^2$) that generates the same 
${\cal Z}\left[M^2,\displaystyle{\frac{\bar g^2}{2\pi}}\right]$:
\begin{equation}
a_\pm={1\over2}\left[M^2\pm\sqrt{M^4-\frac{4M^2}{1+{\bar g^2/2\pi}}}
\right],\ \ \ \ r_\pm^2=1-\frac{\bar g^2}{2\pi}a_\pm.
\end{equation}
We have of course to require that they correspond to real values of 
$a_\pm$ and $r_\pm^2$, and that the unitarity condition ({\it i.e.} 
$a>r^2$) is respected. Actually the equivalent choices of the initial 
parameters are potentially more because from the expression of the partition 
function we see that
$${\cal Z}\left[M^2,\frac{\bar g^2}{2\pi}\right]=
{\cal Z}\left[M^2,-
\frac{{\bar g^2/2\pi}}{1+{\bar g^2/2\pi}}\right].
$$
One can show \cite{Frie} that this duality symmetry is related to the choice 
of compactification radius $R$ in the bosonized version of the theory, namely
$R=\left(1+\displaystyle{\frac{\bar g^2}{2\pi}}\right)^{{1\over 2}}$ or 
$R=\left(1+\displaystyle{\frac{\bar g^2}{2\pi}}\right)^{-{1\over 2}}$. This
property allows 
us to limit our study to $\displaystyle{\frac{\bar g^2}{2\pi}}>0$, the
partition function 
being the same for the choices
\begin{equation}
\hat a_\pm={1\over2}\left[M^2\pm\sqrt{M^4-4 M^2
\left (1+\displaystyle{\frac{\bar g^2}{2\pi}}\right)}
\right],\ \ \ \ 
\hat r_\pm^2=1+\frac{{\bar g^2/2\pi}}{1+{\bar g^2/2\pi}}\hat{a}_\pm.
\end{equation}
We skip the details of the analysis giving the complete list of the possible 
choices (we take $e^2/4\pi=1$ from now on):\hfill\\
\begin{center}
\begin{tabular}{||c |c | l||}\hline
       &~~ $\displaystyle{{4}\left(1+\gpip\right)^{-1}}<M^2<4\left
(1+\gpip\right)$~~
       &~~2 solutions: $a_\pm$, $r^2_\pm$\\
$0<\displaystyle{{\bar g^2}}<\frac{\pi}{2}$&
~~$4\left (1+\gpip\right)<M^2< {1+2\pi/\bar g^2}$~~&
~~4 solutions: $a_\pm$, $r^2_\pm$, $\hat a_\pm$, $\hat r^2_\pm$\\
&~~$M^2>{1+2\pi/\bar g^2}$~~&~~3 solutions: $a_\pm$, $r^2_\pm$, $\hat a_-$,
$\hat r^2_-$ \\
 \cline{1-3}
   &$\displaystyle{4 \left(1+\gpip\right)^{-1}}<M^2<{1+2\pi/\bar g^2}$~  &
~~2 solutions:  $a_\pm$, $r^2_\pm$\\
$\displaystyle{\frac{\pi}{2}}<\displaystyle{{\bar g^2}}<2\pi$ &
${1+2\pi/\bar g^2}<M^2<4\left (1+\gpip\right)$
 & ~~1 solution: $a_-$, $r_-$ \\
   &$M^2>4\left (1+\gpip\right)$
 & ~~3 solutions: $a_\pm$, $r^2_\pm$, $\hat a_-$, $\hat r^2_-$  \\
   \cline{1-3}
&~~  $1+2\pi/\bar g^2<M^2<4\left (1+\gpip\right)$~~& ~~1 solution: $a_-$,
$r_-$ \\
$\displaystyle{{\bar g^2}>{2\pi}}$
&~~$M^2>4\left (1+\gpip\right)$ & ~~3 solutions: $a_\pm$, $r^2_\pm$, $\hat
a_-$, $\hat r^2_-$\\
\hline
\end{tabular}
\end{center}
No solution is possible for
$M^2<4(1+\bar g^2/2\pi)^{-1}$ and
$\displaystyle{\frac{\bar g^2}{2\pi}}<1$, or $M^2<(1+2\pi/\bar g^2)$ and
$\displaystyle{\frac{\bar g^2}{2\pi}>1}$. We learn that 
a lower bound for the mass exists depending on the strength of the Thirring 
coupling and the absolute minimum is easily seen to be $M^2=1$ 
(for $\displaystyle{\frac{\bar g^2}{2\pi}}\rightarrow \infty$). 
In the region where only two values are allowed a simple duality 
transformation, $\displaystyle{a'=\frac{a}{a-r^2}\,\,\,\,\,\,\, {r'}^2
=\frac{a-1}{a-r^2}}$, connecting them can be found: it relates 
large $a$ to $a, r^2 \simeq 1$. Its self-dual points lie on the curve 
$\displaystyle{a=1+r^2}$ and correspond to 
the vanishing of the derivative of the mass with respect to 
the Jackiw-Rajaraman parameter at fixed $\displaystyle{\frac{\bar g^2}{2\pi}}$. 
This curve, describing the critical line 
$\displaystyle{M^2={4\over 1+\bar g^2/2\pi}}$ 
that bounds the second region $(\displaystyle{\frac{1}{4}<\frac{\bar g^2}
{2\pi}<1})$ from below, is the generalization 
of $a=2$ point of the chiral case. 
The fact that $a=2$ must have some relevance, in the space of two 
dimensional chiral gauge theories, has often been claimed in the 
literature \cite{Michi}, even at non abelian level where a change in 
the constraints structure has been noticed \cite{Due}. 
In non abelian anomalous gauge models 
the integration over the gauge field cannot be generally performed and 
few exact results are known. However the bosonized pure chiral case 
(in the gauge invariant formulation \cite{Hara}) was studied at perturbative 
level by Oz \cite{Oz} in a covariant gauge and by us \cite{Us} in light-cone 
gauge. At one loop level the parameter $a$ acquires a dependence 
from a renormalization scale, due to the ultraviolet divergencies of the 
theory. The relevant one-loop $\beta-$function has a fixed point exactly at 
$a=2$, and unitarity appears to be preserved for $a>1$. In view of our 
abelian analysis is tempting to notice that if the previous exact duality 
carries over to the non abelian case, it entails the vanishing 
of the $\beta-$function at its self-dual point $\displaystyle{a=2}$. 
It would be interesting to test, at least at 
perturbative level, our conjecture in non abelian models more general 
than the pure chiral case. Unfortunately the interpolating situation 
cannot be generalized to a non abelian symmetry, using only one gauge 
field, as one easily realizes. We introduce therefore a different 
abelian theory, that is more suitable to a non abelian generalization (we set 
the bare Thirring coupling $g^2=0$)
\begin{equation}
\hat{\cal L}=\sum_{i=1}^{N_+}\bar\psi_i 
i\gamma^\mu(\nabla_\mu+i({1+\gamma_5\over 2})A_\mu)\psi_i 
+\sum_{i=1}^{N_-}\bar\psi_i i\gamma^\mu(\nabla_\mu+i({1-\gamma_5\over 
2})A_\mu)\psi_i +{1\over 4e^2}F^{\mu\nu}F_{\mu\nu},
\end{equation}
describing $N_+$ right and $N_-$ left favours, both interacting 
with the same electromagnetic field. [A similar model with a different 
number of left and right movers has been discussed in \cite{Kiku}. There, 
the potential problems have been avoided by cancelling the anomaly 
with suitable choice of left and right electric charges, in analogy with the 
Standard Model.] Introducing the Hodge decomposition
$\displaystyle{
A_\mu={2\pi\over L}a_\mu+\partial_\mu\phi_1+\eta^\mu_\nu\partial_\nu\phi_2,}
$
we can compute along the previous line the local part of the partition 
function (the Jackiw-Rajaraman parameter is taken equal for all the flavours)
\begin{equation}
\hat{\cal
Z}_{Loc.}=\exp\left[-(N_++N_-)S_{Liou.}\right]\det\left(-{\Delta\over
\hat M^2}+1\right)^{-{1\over 2}}\sqrt{{a^2(N_++N_-)^2-4N_+N_-\over 4}},
\end{equation}
where $\hat{e}^2=(N_++N_-)e^2$ and the mass is
$\displaystyle{
\hat{M}^2={\hat{e}^2\over4\pi(a-1)}\left (a^2-{4N_+N_-\over (N_++N_-)^2}
\right)}.$
The determinants involved in the global part leads to
\begin{eqnarray}
\hat{\cal Z}_{Glob.}=&&\Theta^{N_-}[0,0](0,\tau)\Theta^{* 
N_+}[0,0](0,\tau)
\int_{-\infty}^{+\infty}da_1 da_2 
\Theta^{N_+}[-a_2,a_1](0,\tau)\Theta^{* N_-}[-a_2,a_1](0,\tau)
\nonumber\\
&&\exp\left[-{\pi\over 2\tau_2}\left((a_1-\tau_1a_2)^2+\tau^2_2 
a_2^2\right)(N_++N_-)(a-1)+2\pi i(N_+-N_-)a_1a_2\right].
\label{Grosso}
\end{eqnarray}
The $a_\mu$-independent factor $\Theta^{* N_+}[0,0](0,\tau)\Theta^{ 
N_-}[0,0](0,\tau)$ corresponds to the (free) fermionic partners, that are 
decoupled from the electromagnetic field. We could omit this term if we 
divide the full partition function by the contribution of $N_+$ left 
fermions and $N_-$ right fermions (only gravitionally coupled). In this 
case we have to substitute in ${\cal Z}_{Loc.}$ the Liouville action by the 
appropriate one from 
$N_+$ right and $N_-$ left fields, where a "gravitational" 
Jackiw-Rajaraman parameter appears due to the Lorentz anomaly 
\cite{Pery92,Basti92}. 
We notice that in the genus one case no globality problems arise because 
the tangent-bundle is trivial (see \cite{Pery92} for a discussion of this 
point). Coming back to the integral in (\ref{Grosso}) we obtain
\begin{equation}
\hat{{\cal Z}}_{Glob.}= \sqrt{{4\over a^2(N_++N_-)^2-4N_+N_-}}
\Theta_{2(N_++N_-)}(0,\hat\Lambda),
\end{equation}
where the $2(N_++N_-)$ dimensional theta function is defined by
$\displaystyle{
\Theta_{2(N_++N_-)}(0,\hat\Lambda)=\!\!\!\!\!\!\!\!\!\!\!
\sum_{n_i,n_j\in 
Z^{2(N_++N_-)}}\!\!\!\!\exp[i\pi n_i \hat\Lambda_{ij}n_j ],
}$
the covariance matrix being
\begin{equation}
\label{covariance2}
\hat{\Lambda}=\left (\begin{array}{cc}
1_{N_+}\tau &  0\\
\\0 & -1_{N_-}\bar{\tau}\end{array} \right )-
i {4\tau_2 \over a^2-{4N_+N_-\over (N_++N_-)^2}}\left (\begin{array}{cc}
{2N_-\over (N_++N_-)^2}1_{N_+N_+}& -{a\over (N_++N_-)}1_{N_+N_-}\\
\\-{a\over (N_++N_-)}1_{N_-N_+} & {2N_+\over (N_++N_-)^2}1_{N_-N_-}
\end{array} \right ).
\end{equation}
Here $1_{N_iN_j}$ means the $N_i\times N_j$ matrix with $1$ in all the 
entries. 
Subtracting the contribution of the electromagnetic-free fermionic partners 
we have (we omit from now on the gravitational part that is presented 
for example in \cite{Pery92})
\begin{equation}
\hat{\cal Z}=\det\left(-{\Delta\over \hat{M}^2}+1\right)^{-{1\over 2}} 
\Theta_{2(N_++N_-)}(0,\hat\Lambda).
\end{equation}
From requiring the positivity of the mass we get
\begin{equation}
a>1\,\,\,\,\,\, {\rm or}\,\,\,\,\,\, a<\sqrt{4N_+N_-\over (N_++N_-)^2},
\end{equation}
but only in the first case the series defining the multidimensional 
theta function converges. We see therefore that in this model the 
unitarity request on a single ({\it e.g.} $N_+=1,N_-=0$) chiral 
fermion ($a>1$) implies the unitarity for generic $N_+$ and $N_-$. We 
have now to discover what is the fermionic model described by the 
$\Theta-$function: let us define the coupling
\begin{equation}
{\hat{g}^2\over 2\pi}={4\over (N_++N_-)(a-1)}.
\end{equation}
It is not difficult to show that the $\Theta-$function is generated 
by $N_+$ right and $N_-$ left fermions interacting through a 
current-current term
\begin{equation}
{\cal L}_{I}={\hat{g}^2\over 2}\sum^{N_+}_{i=1} 
\sum^{N_-}_{j=1}J_+^{i\mu}J_{-\mu}^j\ \ \ \
J_{i\pm}^\mu=i\bar{\psi}^i\gamma^\mu({1\pm\gamma_5\over 2})\psi^i.
\end{equation}
For $N_+=N_-=1$ we recover exactly a Thirring model. We see that no 
"dual" behaviour is apparently present in the general case, even if the mass 
of the emerging boson does: the fermionic part of the partition function does 
not depend only on the coupling ${\hat{g}^2\over 2\pi}$ but on the number of 
flavours too. It could be that a more sophisticated analysis 
is needed in order to make manifest duality properties. In any case 
this does not mean that the non abelian case must not 
posses some interesting duality (or self-duality) properties: in the 
abelian case the conformal invariance of the fermionic sector is always 
achieved while we expect, in the non abelian situation, its recovering 
only at particular points in the parameter space, as the Wess-Zumino 
action does. A non abelian analysis deserves therefore a closer look and 
we expect a non-trivial renormalization group behaviour for the 
anomalous theory. The problem is currently under scrutiny.

\section*{Acknowledgement}

We would like to thank Prof. R. Jackiw a critical reading of the manuscript 
and Prof. H. Neuberger for pointing out ref. \cite{Adamo,Kiku}.
 
\medskip

This work is supported by NSF grant PHY-9315811, 
in part by funds provided by the U.S. D.O.E.
under cooperative agreement \#DE-FC02-94ER40818
and by INFN, Frascati, Italy. 

\end{document}